# Effect of quantum interference on the optical properties of a three-level V-type atomic system beyond the two-photon resonance condition

S M Mousavi[1], L Safari[1], M Mahmoudi[1] and M Sahrai[2]

[1] Physics Department, Zanjan University, PO Box 45195-313, Zanjan, Iran
[2] Research Institute for Applied Physics and Astronomy, University of Tabriz, Tabriz, Iran

**Abstract**
The effect of quantum interference on the optical properties of a pumped-probe three-level V-type atomic system is investigated. The probe absorption, dispersion, group index and optical bistability beyond the two-photon resonance condition are discussed. It is found that the optical properties of a medium in the frequency of the probe field, in general, are phase independent. The phase dependence arises from a scattering of the coupling field into the probe field at a frequency which in general differs from the probe field frequency. It is demonstrated that beyond the two-photon resonance condition the phase sensitivity of the medium will disappear.

## 1. Introduction

Coherent or incoherent interaction between electromagnetic waves and atoms induces coherence among atomic states, which leads to interesting quantum interference effects [1]. Quantum interference and quantum coherence in an atomic system can lead to many important optical phenomena such as lasing without inversion [2], enhanced index of refraction [3], electromagnetically induced transparency [4], superluminal light propagation [5] and optical bistability (OB) [6]. Atomic coherence can be reduced via decay processes, such as optical pumping, collision, laser intensity and laser field coherence. The relative phase between applied laser fields in an atomic system is an important parameter for controlling the atomic coherence. It is well known that the optical properties of a closed atomic system interacting with laser fields are completely phase dependent [7–11]. Recently, it was shown that the phase-dependent behaviour, in a closed-loop system, is valid only under the multi-photon resonance condition, so the phase-dependent process contributing to the probe field susceptibility occurs only at a specific frequency [12].

The effects of quantum interference on a variety of optical effects have been theoretically and experimentally reviewed [13, 14]. Interference in stimulated emissions was first reported in the experiments by Garrett [15]. Phase control of quantum interference effects was provided in the experiments by Koerunsky *et al* [16]. The phase-dependent behaviour in atomic systems can also be induced by quantum interference due to spontaneous emission [17, 18]. Gain components in an Autler–Townes doublet due to quantum interference in a pumped-probe three-level V-type system was investigated [19]. The effect of such interference on OB [20, 21] and group velocity [22] has also been discussed.

The effect of the relative phase between applied fields with spontaneously generated coherence (SGC) [23] on probe gain with and without population inversion has also been investigated [24]. It is shown that the probe gain can be changed just by tuning the relative phase between the probe and pump fields. In another study, transient response of a pumped-probe three-level V-type atomic system in the presence of quantum interference was discussed [25]. It was demonstrated that the transient response of the system can almost be eliminated just by choosing the proper relative phase between the two applied fields. The effect of the relative phase on transient and steady-state behaviours of a four-level atomic medium in a closed-loop configuration has been discussed [26, 27]. Taking an overview of many proposals, we note that the two-photon resonance condition has been employed to obtain the phase-dependent behaviour of the systems.

In this paper, we investigate the effect of quantum interference on the optical properties of a pumped-probe three-level V-type atomic system beyond the two-photon resonance condition. In particular, we study the probe absorption, dispersion, group index and OB in this system. We apply the Floquet decomposition to the equation of motion to solve the time-dependent differential equations. This leads to identification of the different scattering processes contributing to the medium response. We find that the various Floquet components can be interpreted in terms of different scattering processes. So, the medium in the frequency of the probe field, in general, is not phase dependent. The phase dependence arises from a scattering of the coupling field into the probe field at a frequency which in general differs from the probe field frequency.

## 2. Theoretical analysis

### 2.1. Model and equations

Consider a closed V-type three-level atomic system with two nearly degenerate excited levels $|2\rangle$ and $|3\rangle$ and a ground level $|1\rangle$, as shown in figure 1(a). A strong coherent field with frequency $\omega_c$ and amplitude $\vec{E}_c$ couples the $|1\rangle \rightarrow |2\rangle$ transition. A weak probe field of frequency $\omega_p$ and amplitude $\vec{E}_p$ couples the $|1\rangle \rightarrow |3\rangle$ transition. Since the dipole moments are not orthogonal, we have to consider an arrangement where each field (pump and probe) acts only on one transition. This can be achieved by considering the case shown in figure 1(b), where the probe (pump) acts on the transition $|1\rangle \rightarrow |3\rangle$ ($|1\rangle \rightarrow |2\rangle$) [19, 22, 28]. Therefore, we assume that the strong coupling field is right-hand polarized ($\sigma^+$), while a weak probe field is left-hand polarized ($\sigma^-$) [5, 23, 29]. The two excited levels are coupled to the ground level by the same vacuum mode. The spontaneous decay rates from the level $|2\rangle$ and the level $|3\rangle$ to the ground level $|1\rangle$ are denoted by $\gamma_2$ and $\gamma_3$, respectively.

The Hamiltonian in interaction picture and under the dipole and rotating wave approximations is given by

$$H = -\hbar\Omega_p \, e^{i\Delta_p t}|3\rangle\langle 1| - \hbar\Omega_c \, e^{i\Delta_c t}|2\rangle\langle 1| + \text{c.c.}, \quad (1)$$

where $\Omega_p = \frac{\vec{E}_p \cdot \vec{d}_{13}}{2\hbar}$ ($\Omega_c = \frac{\vec{E}_c \cdot \vec{d}_{12}}{2\hbar}$) is the Rabi frequency of the probe (coupling field). $\Delta_c = \omega_c - \omega_{21}$ and $\Delta_p = \omega_p - \omega_{31}$ are the detunings between the applied fields and the corresponding atomic transitions, where $\hbar\omega_{ij}$ corresponds to the energy difference between the levels $i$ and $j$.

The master equation of motion for the density operator in an arbitrary multilevel atomic system can be written as

$$\frac{\partial \rho}{\partial t} = \frac{1}{i\hbar}[H, \rho] + L\rho, \quad (2)$$

$L\rho$ represents the decay part in the system. By expanding equation (2), we can easily arrive at the density matrix equation

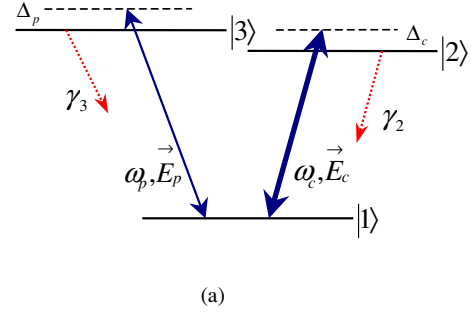

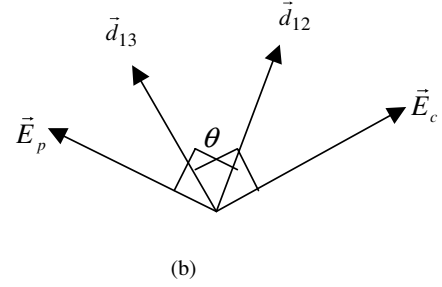

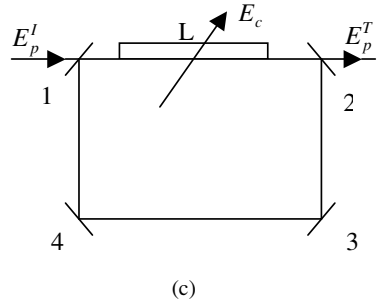

**Figure 1.** (a) Schematic energy diagram of a three-level V-type atomic system driven by probe and coupling fields. (b) The polarization is chosen such that one field only drives one transition. (c) Unidirectional ring cavity with atomic sample of length $L$. $E_P^I$ and $E_P^T$ are the incident and transmitted fields, while $\vec{E}_p$ and $\vec{E}_0$ are the probe and coupling fields, respectively. For mirrors 1 and 2 it is assumed that $\bar{R} + \bar{T} = 1$, and mirrors 3 and 4 have perfect reflectivity.

of motions:

$$\begin{aligned}
\dot{\rho}_{22} &= -2\gamma_2\rho_{22} + i\Omega_c\rho_{12} - i\Omega_c\rho_{21} - \eta\sqrt{\gamma_3\gamma_2}\,(\rho_{23} + \rho_{32}) \\
\dot{\rho}_{33} &= -2\gamma_3\rho_{33} + i\Omega_p\,e^{-i\delta t}\rho_{13} \\
&\quad - i\Omega_p\,e^{i\delta t}\rho_{31} - \eta\sqrt{\gamma_3\gamma_2}\,(\rho_{23} + \rho_{32}) \\
\dot{\rho}_{12} &= -(\gamma_2 + i\Delta_c)\rho_{12} \\
&\quad + i\Omega_c(\rho_{22} - \rho_{11}) + i\Omega_p\,e^{i\delta t}\rho_{32} - \eta\sqrt{\gamma_3\gamma_2}\,\rho_{13} \\
\dot{\rho}_{13} &= -(\gamma_3 - i(\delta - \Delta_p))\rho_{13} \\
&\quad + i\Omega_p\,e^{i\delta t}(\rho_{33} - \rho_{11}) + i\Omega_c\rho_{23} - \eta\sqrt{\gamma_3\gamma_2}\,\rho_{12} \\
\dot{\rho}_{23} &= -[(\gamma_3 + \gamma_2) - i(\Delta_c - \Delta_p + \delta)]\rho_{23} + i\Omega_c\rho_{13} \\
&\quad - i\Omega_p\,e^{i\delta t}\rho_{21} - \eta\sqrt{\gamma_3\gamma_2}\,(\rho_{22} + \rho_{33}) \\
\rho_{11} &+ \rho_{22} + \rho_{33} = 1,
\end{aligned} \quad (3)$$

where $\delta = \Delta_p - \Delta_c$ is the two-photon resonance detuning, i.e. the pump-probe detuning. Moreover, we assume $\omega_{31} \cong \omega_{21}$.

The terms including $\eta\sqrt{\gamma_3\gamma_2}$ represent the effect of quantum interference among decay channels where $\eta = \frac{\vec{d}_{12}\cdot\vec{d}_{31}}{|\vec{d}_{21}||\vec{d}_{31}|} = \cos\theta$ represents the strength of the interference in spontaneous emission. It depends on the angle between the two dipole moments $\vec{d}_{21}$ and $\vec{d}_{31}$. When the two dipole moments are parallel the effect of quantum interference is maximum and $\eta = 1$, whereas for the orthogonal dipole moments there is no interference due to spontaneous emission and $\eta = 0$. Note that for nearly degenerate upper levels, i.e. $\omega_{31} \cong \omega_{21}$, this coherence becomes important, but for large upper level spacing the effect of quantum interference may be dropped [23, 24].

### 2.2. Linear susceptibility and group velocity

The response of the atomic system to the applied fields is determined by the susceptibility $\chi$, which is defined as [30]

$$\chi(\omega_p) = \frac{2Nd_{31}}{\varepsilon_0 E_p}\rho_{31}(\omega_p), \quad (4)$$

where $N$ is the atom number density in the medium. The real and imaginary parts of $\chi$ correspond to the dispersion and the absorption of a weak probe field, respectively. For further discussion, we introduce the group index $n_g = \frac{c}{v_g}$ where $c$ is the speed of light in vacuum and the group velocity $v_g$ is given as

$$v_g = \frac{c}{1 + 2\pi'\chi(\omega_p) + 2\pi\omega_p\frac{\partial}{\partial\omega_p}\chi'(\omega_p)}, \quad (5)$$

where $\chi'(\omega_p)$ is the real part of the susceptibility. The group velocity of a light pulse can be determined by the slope of the dispersion. In a dispersive medium, the frequency components of a light pulse experience different refractive indices, and the group velocity of a light pulse in such a material can exceed the speed of light in vacuum, leading to superluminal light propagation. In our notation, the negative slope of dispersion corresponds to superluminal light propagation, while the positive slope shows subluminal light propagation. In addition, negative (positive) values in the imaginary part of susceptibility show the gain (absorption) for the probe field.

### 2.3. Optical bistability

Optical bistability (OB) is a result of the nonlinearity of the interactivity atomic medium and the feedback of the optical interactivity field from the cavity mirrors. The bistable behaviour of the above-described atomic system will be investigated in the optical ring cavity as shown in figure 1(c). For simplicity, we assume that mirrors 3 and 4 have 100% reflectivity, and the intensity reflection and transmission coefficients of mirrors 1 and 2 are $\bar{R}$ and $\bar{T}$ (with $\bar{R} + \bar{T} = 1$). The three-level atomic system, whose dynamics is described by equations (3), is a collection of $N$ homogeneously broadened atoms contained in a cell of length $L$. The total electromagnetic field seen by these atoms is $\vec{E} = \vec{E}_p e^{-i\omega_p t} + \vec{E}_c e^{-i\omega_c t} + \text{c.c.}$, where the probe field circulates in the ring cavity and the pumping field does not circulate in the cavity. Then under slowly varying envelope approximation, the dynamic response of the probe field is governed by Maxwell's equation [30]

$$\frac{\partial E_p}{\partial t} + c\frac{\partial E_p}{\partial z} = i\frac{\omega_p}{2\varepsilon_0}P(\omega_p), \quad (6)$$

where $\varepsilon_0$ is the permittivity of free space. $P(\omega_p)$ is the induced polarization in the transition $|1\rangle \to |3\rangle$ and is given by $P(\omega_p) = N\mu_{31}\rho_{31}$. For a perfectly tuned ring cavity, in the steady-state case, the boundary conditions between the incident field $E_p^I$ and the transmitted field $E_p^T$ lead to

$$E_p(L) = E_p^T/\sqrt{\bar{T}}, \qquad E_p(0) = \sqrt{\bar{T}}E_p^I + \bar{R}E_p(L). \quad (7)$$

Note that the feedback mechanism of the probe field due to the mirrors for the nonlinear atomic medium is responsible for the behaviour of OB. It means that for $\bar{R} = 0$, no bistability can occur. In the mean-field limit by using the boundary conditions, i.e. equations (7), the steady state of the transmitted field is given by

$$y = x - 2iC\gamma_3\rho_{31}(x), \quad (8)$$

where $x = \frac{\mu_{31}E_p^T}{\hbar\sqrt{\bar{T}}}$ and $y = \frac{\mu_{31}E_p^I}{\hbar\sqrt{\bar{T}}}$, and $C = \frac{N\omega_p L\mu_{31}^2}{2\hbar\varepsilon_0 c\bar{T}\gamma_3}$ is the usual cooperation parameter. Equation (8) shows that the probe coherence term has an important role in establishing the OB behaviour.

### 2.4. Two-photon resonance condition

We assume that the two-photon resonance condition, i.e. $\delta = 0$, is to be fulfilled. Then the coefficients of equations (3) do not have explicit time-dependent terms. Under the weak probe field approximation and for $\Delta_p = \Delta_c = 0$ and $\gamma_3 = \gamma_2 = 1$, a simple analytical expression for the coherence term $\rho_{31}$ can be found as follows:

$$\rho_{31} = \frac{-4i\eta\Omega_c^5 + 16i\eta\Omega_c(\eta^2 - 1)^2}{D}$$
$$+ \frac{-16i(\eta^2-1)^2 - 4i\Omega_c^2(4-5\eta^2+3\eta^4) + 2i\Omega_c^4(\eta^2-2)}{D}\Omega_p$$
$$+ \frac{-4i\Omega_c^2\eta^2(3\eta^2-5) + 2i\Omega_c^4\eta^2}{D}\Omega_p^*, \quad (9)$$

where

$$D = -16 + 48\eta^2(1-\eta^2)$$
$$+ 16\Omega_c^2(\eta^4 + 2\eta^2 - 3) + 4\Omega_c^4(\eta^2 - 9) - 8\Omega_c^6.$$

The three terms expressed in equation (9) involve different physical processes which have simple interpretations (see figure 2) similar to [12]. The first part of equation (9) is proportional to $\eta\Omega_c$ and represents the scattering of the coupling field into the probe field mode via the quantum interference due to spontaneous emission and depends on the phase difference between the two fields. The coupling field excites the electron from the level $|1\rangle$ to the level $|2\rangle$. The overlapping of the vacuum modes of two spontaneous decays sends the electron to the level $|3\rangle$, and then it contributes to the probe susceptibility. In the absence of the quantum interference due to spontaneous emission, this term does not contribute to the probe susceptibility, and then the response of the system to the probe field is not phase dependent. The second term in equation (9), which is proportional to $\Omega_p$, represents the direct response of the medium to the probe field at the probe field frequency. It involves the excitation and dexcitation of the probe transition, which does not depend on the relative phase. The last term, which is proportional to $\Omega_p^*$, shows a counter-rotating contribution. This preliminary interpretation of the individual contributions will become clearer in the time-dependent analysis beyond the two-photon resonance condition.

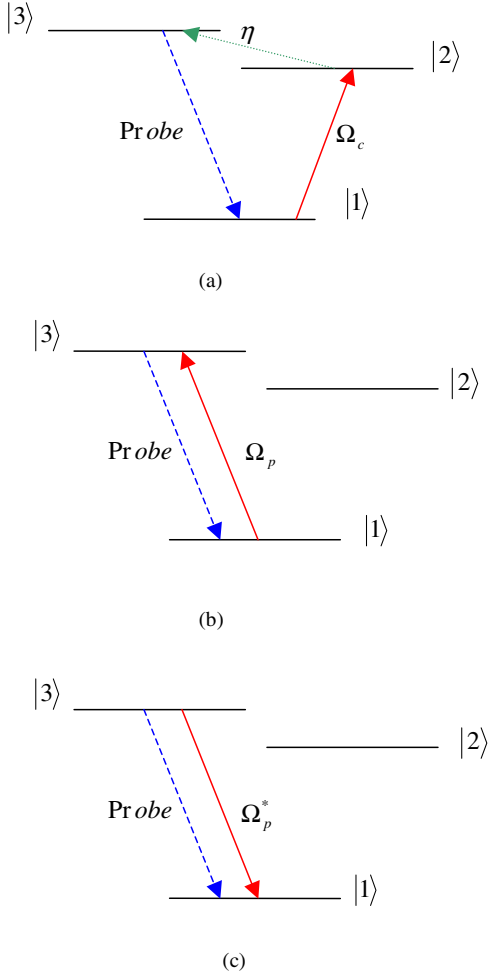

**Figure 2.** Interpretation of the different contributions to the probe field susceptibility in terms of transition pathway. Here, (a) represents the interaction of the coupling field with quantum system and quantum interference due to spontaneous emission, (b) is the direct scattering of the probe field in the probe transition and (c) shows a counter-rotating term. The solid red arrows indicate the coupling field and the dashed blue arrows are probe field interactions. The effect of quantum interference due to the spontaneous emission is shown by a dotted green arrow.
(This figure is in colour only in the electronic version)

By considering $\Omega_c = |\Omega_c| e^{i\varphi_c}$, $\Omega_p = |\Omega_p| e^{i\varphi_p}$ and redefining the new atomic density matrix variables in equation (1), we obtain equations for the redefined density matrix elements which are identical with equation (1); only the parameter $\eta$ is replaced by $\eta e^{i\Delta\varphi}$ where $\Delta\phi = \varphi_c - \varphi_p$. Then the first term in equation (9), which has odd powers of $\eta$, becomes phase dependent, while the second and third terms are not phase dependent because of the even powers of $\eta$. It means that by the change in the phase difference between the applied fields, the relative phase of the dipole moments will be changed, and thus it changes the parameter $\eta$.

*2.5. Beyond the two-photon resonance condition*

We now evaluate the equation of motion, equations (3), for the general case of time-dependent coefficients. These equations can be written in a compact form as

$$\frac{\partial R}{\partial t} + \Sigma = MR. \quad (10)$$

Here

$$R = (\rho_{11}, \rho_{22}, \rho_{12}, \rho_{21}, \rho_{31}, \rho_{13}, \rho_{23}, \rho_{32})^T \quad (11)$$

is a vector containing the density matrix elements and

$$\Sigma = (-2\gamma_3, 0, -i\Omega_c, i\Omega_c, 0, 0, \eta, \eta)^T \quad (12)$$

is a vector, independent of the density matrix elements. According to the Floquet decomposition, both matrices $M$ and $\Sigma$ can be separated into terms with different time dependences [31, 32],

$$M = M_0 + \Omega_p M_1 e^{-i\delta t} + \Omega_p^* M_{-1} e^{i\delta t}, \quad (13)$$

where $\Sigma_0$, $\Sigma_{\pm 1}$, $M_0$ and $M_{\pm 1}$ are time-independent coefficients. A simple application of Floquet's theorem to equations (3) shows that the stationary solution $R$ will have only the terms at the harmonics of the detuning $\delta$. Then the matrix $R$ can be written as

$$R = R_0 + \Omega_p R_1 e^{-i\delta t} + \Omega_p^* R_{-1} e^{i\delta t} + \text{higher order}. \quad (14)$$

For a weak probe field interaction, the terms at higher order harmonics of equation (14) are negligible. Then, using equations (13) and (14) in equation (10), $R_0$ and $R_{\pm 1}$ are given by

$$R_0 = M_0^{-1} \Sigma_0, \quad (15a)$$
$$R_1 = (M_0 + i\delta)^{-1}(\Sigma_1 - M_1 R_0), \quad (15b)$$
$$R_{-1} = (M_0 - i\delta)^{-1}(\Sigma_{-1} - M_{-1} R_0). \quad (15c)$$

In the calculation of the absorption and dispersion as well as the group velocity, we assume a weak probe field interaction, and only the terms at first-order harmonics of equation (14) are used, while in calculation of the OB the contribution of higher-order harmonics become important. The response of the medium is determined by the fifth component of $R$. It can be seen that the contribution proportional to $R_1$ oscillates in phase with the probe beam, and thus contributes to the probe beam susceptibility independent of the frequency of the incident coupling field. For $\delta \neq 0$, two other contributions proportional to $R_0$ and $R_{-1}$ oscillate at different frequencies, and thus do not contribute to the probe beam susceptibility. Even in the limit $\delta \to 0$, the contribution of $R_1$ is distinct from the other two contributions that can propagate in different wave vector directions. Therefore, in general, only the contribution proportional to $R_1$ should be taken into account for calculation of the optical properties of the system such as group velocity and OB.

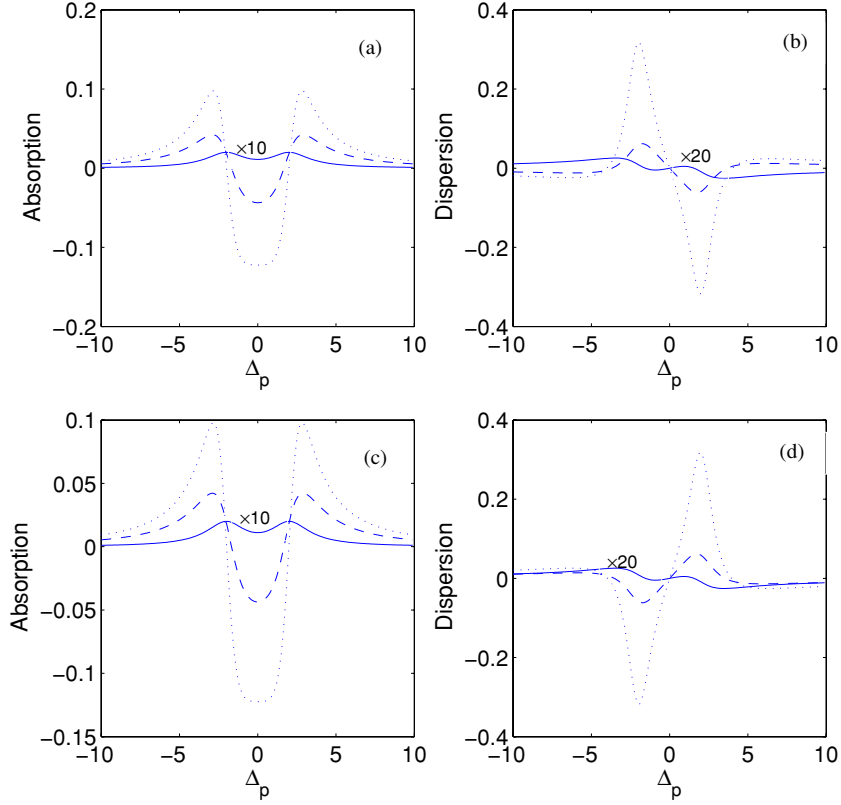

**Figure 3.** The probe absorption (a), (c) and dispersion (b), (d) for $\Delta\varphi = 0$ (a), (b) and $\Delta\varphi = \pi$ (c), (d) versus the probe field detuning. The selected parameters are $\gamma_3 = \gamma_2 = \gamma$, $\Omega_c = 2\gamma$, $\Delta_c = 0$, $\Omega_p = 0.01\gamma$ and $\eta = 0$ (solid), 0.5 (dashed), 0.99 (dotted).

## 3. Results and discussion

We now turn towards the numerical results of the master equations (3). First, we obtain the results near the two-photon resonance condition, i.e. $\delta = 0$. Figure 3 shows the absorption (a, c) and the dispersion (b, d) behaviour of the probe field versus the probe field detuning for different values of the relative phase of the applied fields, i.e. $\Delta\varphi = 0, \pi$. The selected parameters are $\gamma_3 = \gamma_2 = \gamma$, $\Omega_c = 2\gamma$, $\Delta_c = 0$, $\Omega_p = 0.01\gamma$, $\eta = 0$ (solid), 0.5 (dashed), 0.99 (dotted), $\Delta\varphi = 0$ (figures 3(a) and (b)), $\Delta\varphi = \pi$ (figures 3(c) and (d)). Investigation of figure 3 shows that in the presence of quantum interference and near the two-photon resonance condition the optical properties of the system are completely phase dependent. Therefore, the slope of dispersion changes just by changing the relative phase between the coupling and probe fields. Note that in the absence of such interference (solid lines), the phase sensitivity disappears. The phase sensitivity of the optical properties of a pumped-probe three-level V-type system with the quantum interference was mentioned by several authors [24, 25].

In figure 4, we display the group index, $\frac{c}{v_g} - 1$, versus probe field detuning for $\Delta\varphi = 0$ (a) and $\Delta\varphi = \pi$ (b). It can be realized that for $\Delta\varphi = 0$, the group index around $\Delta_p = 0$ is negative, corresponding to superluminal light propagation, while for $\Delta\varphi = \pi$, the group index becomes positive, corresponding to subluminal light propagation.

The similar results shown in figures 3 and 4 are also obtained from the Floquet decomposition. The phase sensitivity in these figures arises from the scattering of the coupling field into the probe field that appears in the probe frequency only under the two-photon resonance condition. According to equation (14), near the two-photon resonance condition, i.e. $\delta = 0$, all the three contributions of $R_0$, $R_1$ and $R_{-1}$ are mixed and then the obtained results are valid only under the two-photon resonance condition. However, if the scattering of the coupling field in the probe field frequency does not have the same propagation direction as the probe field, only the contribution of $R_1$ should be taken into account. For this reason, in figure 5 we consider only the contribution of $R_1$ for displaying the absorption (a), dispersion (b) and group index (c) for $\eta = 0$ (solid), $\eta = 0.5$ (dashed) and $\eta = 0.99$ (dotted). We observe that the slope of dispersion cannot change with the relative phase between the applied fields.

The behaviour of OB near the two-photon resonance condition is displayed in figure 6. The parameters are $\gamma_3 = \gamma_2 = \gamma$, $\Omega_c = 10\gamma$, $\Delta_c = 4.1\gamma$, $\Delta_p = 0$, $\delta = 0$, $C = 400$, $\eta = 0$ (solid), 0.5 (dashed), 0.99 (dotted) as in [20]. In parts (a) and (b), we use all of the three contributions of $R$ for calculating the OB diagram, while in part (c) we choose only the contribution of $R_1$. According to figure 6(a) the quantum interference reduces the threshold of OB. This is in good agreement with the result of [16]. From figures 6(a) and (b) we understand that the behaviour of OB depends on the relative phase between the probe and the coupling fields. Moreover, figure 6(c) implies that by choosing only the contribution of $R_1$ the quantum interference due to spontaneous emission increases the threshold of OB.

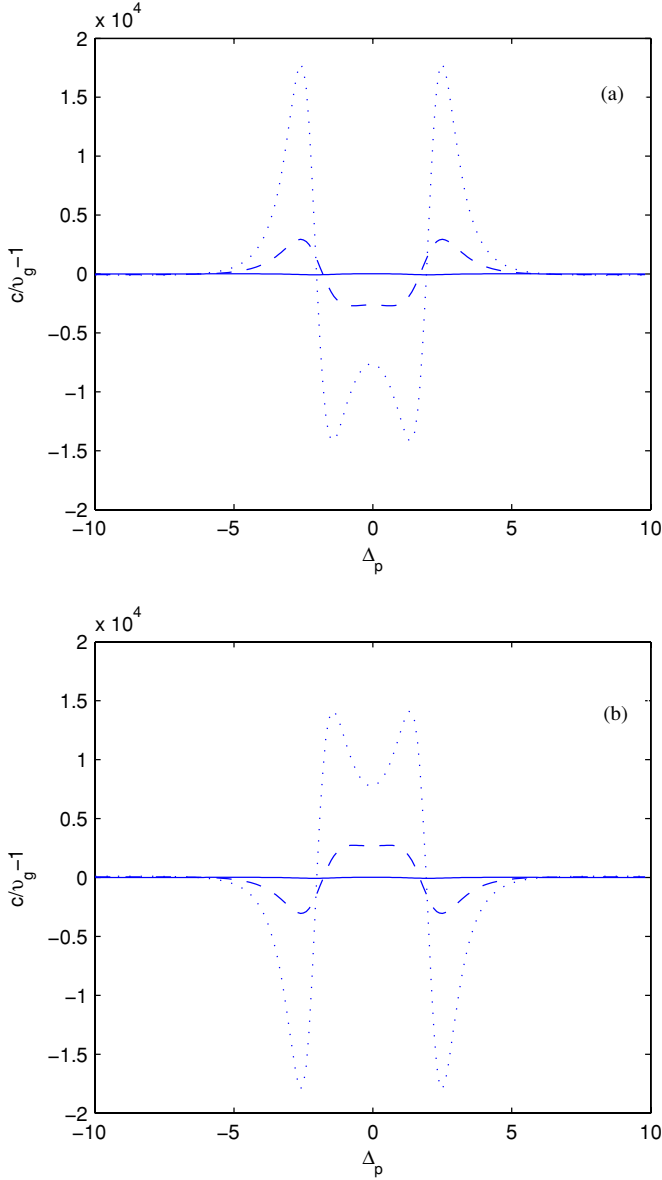

**Figure 4.** Group velocity versus the probe field detuning for $\Delta\varphi = 0$ (a) and $\Delta\varphi = \pi$ (b). The other parameters are the same as in figure 3.

We are interested in calculating the optical properties of the atomic medium beyond the two-photon resonance condition, i.e. $\delta \neq 0$. In figure 7, we show the absorption (a), the dispersion (b) and the group index (c) beyond the two-photon resonance condition. The parameters are same as in figure 3. We take into account the contribution of higher order harmonics (up to six-order harmonics) in the calculation of the probe field contribution. In figure 8, we have plotted the OB in the presence of quantum interference beyond the two-photon resonance condition. The other parameters are the same as in figure 6. It can be seen that the medium is phase independent beyond the two-photon resonance condition.

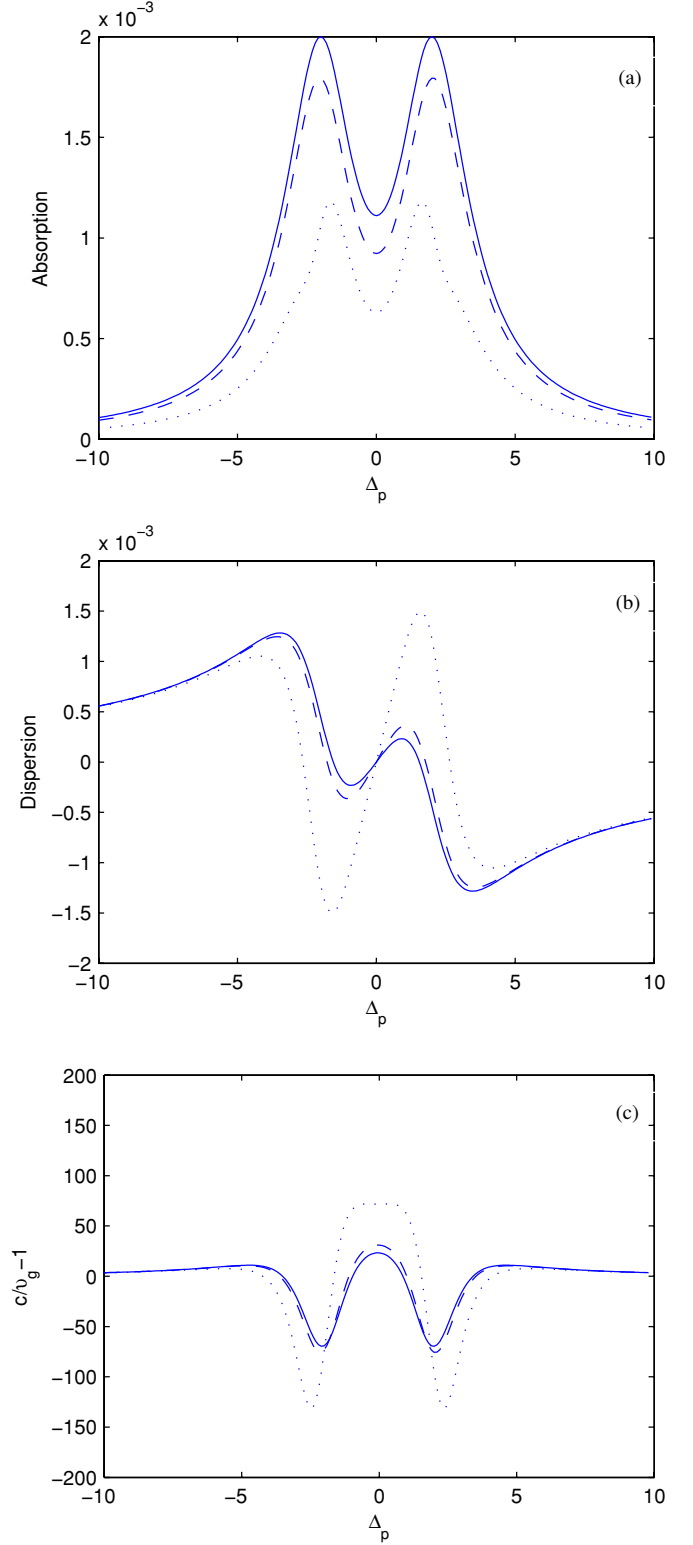

**Figure 5.** The probe absorption (a), dispersion (b) and group index (c) versus probe field detuning. Only the contribution of $R_1$ has been considered. The other parameters are the same as in figure 3.

Physically, beyond the two-photon resonance condition and in the presence of quantum interference, the scattering of the coupling field does not occur at the probe frequency. Thus the output with probe frequency does not show the phase-sensitive behaviour. So, the phase sensitivity in a pumped-probe three-level V-type system only appears under the two-photon resonance condition in which all contributions of the responses of the medium oscillate at the probe frequency.

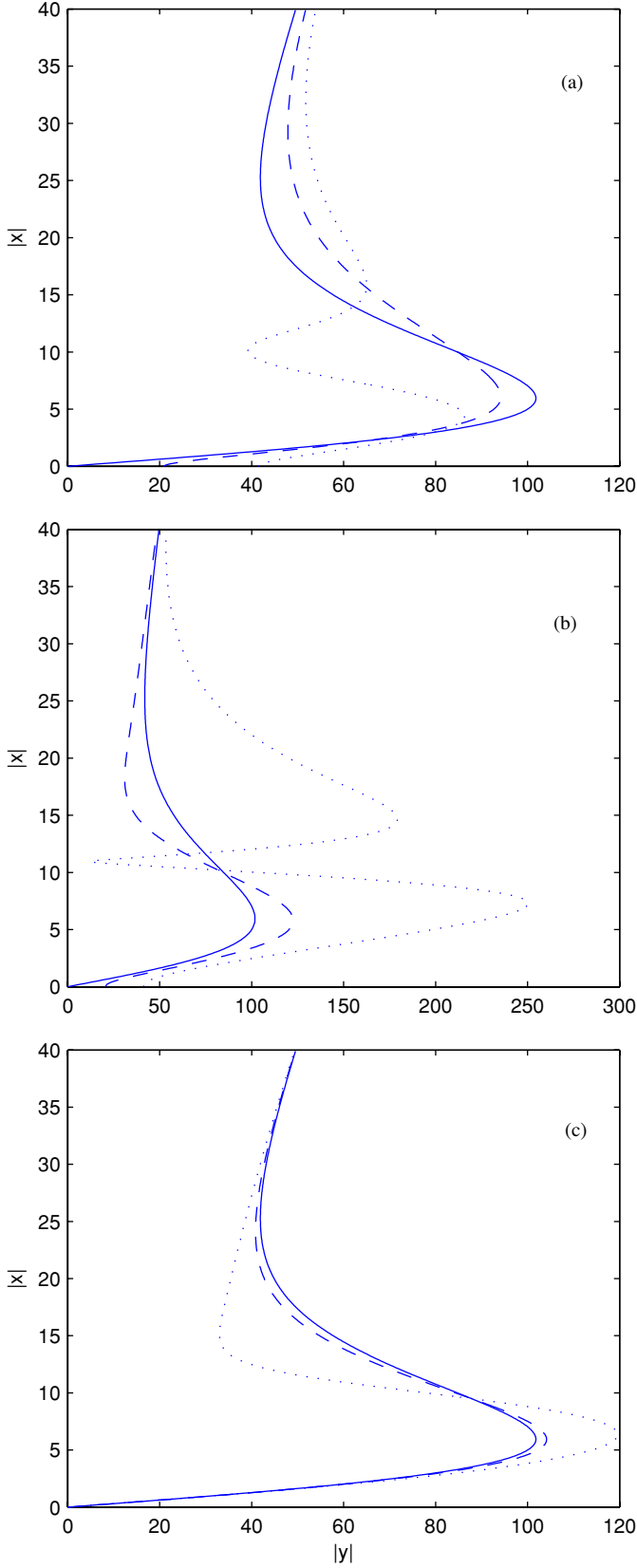

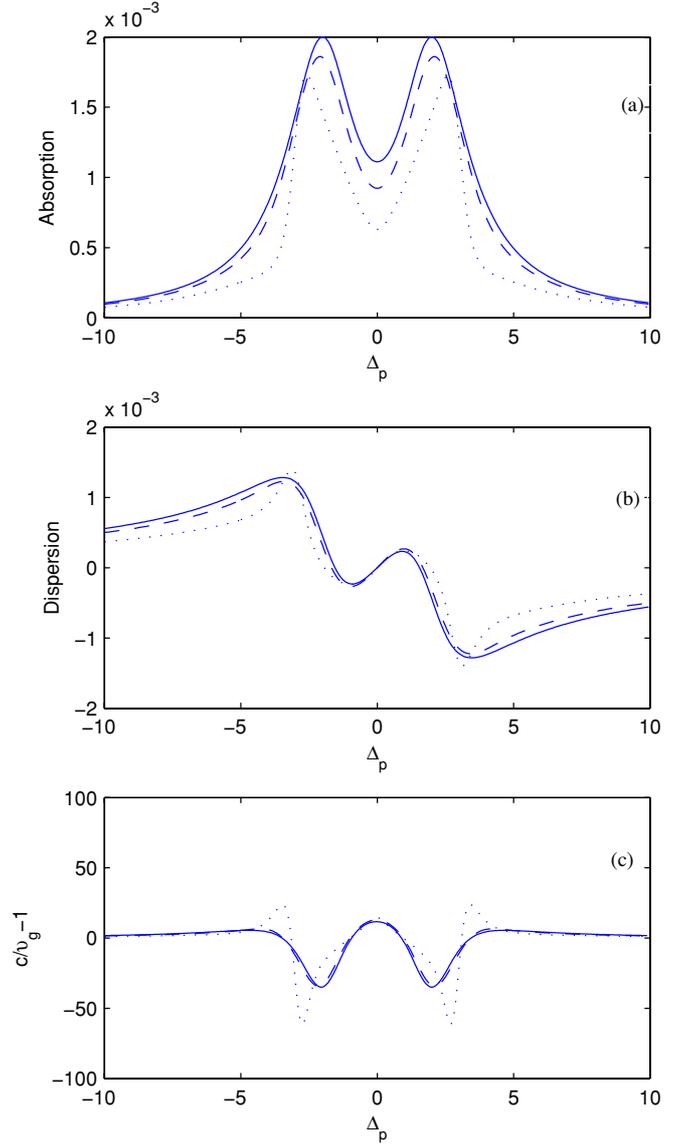

**Figure 6.** The behaviour of OB under the two-photon resonance condition using all contributions of scattering in optical coherence (a) and (b), and only the probe field contribution (c) for $\Delta\varphi = 0$ (a) and $\Delta\varphi = \pi$ (b). The selected parameters are $\gamma_3 = \gamma_2 = \gamma$, $\Omega_c = 10\gamma$, $\Delta_p = 0$, $\Delta_c = -4.1\gamma$, $\delta = 0$, $C = 400$, and $\eta = 0$ (solid), 0.5 (dashed), 0.99 (dotted).

**Figure 7.** Absorption (a), dispersion (b) and group index (c) beyond the two-photon resonance condition. The selected parameters are $\gamma_3 = \gamma_2 = \gamma$, $\Omega_c = 2\gamma$, $\Delta_c = 0$, $\Omega_p = 0.01\gamma$, $\eta = 0$ (solid), 0.5 (dashed), 0.95 (dotted), $\Delta\varphi = 0$ (or $\Delta\varphi = \pi$) and $\delta = \Delta_p$.

In the schematics of a unidirectional ring cavity with four mirrors in OB, the probe field should circulate in the cavity, while the coupling field should not circulate. Therefore, the coupling and probe fields may have different propagation directions, so experimentally it is difficult to establish the OB setup in a pumped-probe three-level V-type system under the two-photon resonance condition. In such a case, one should use the results for beyond the two-photon resonance condition.

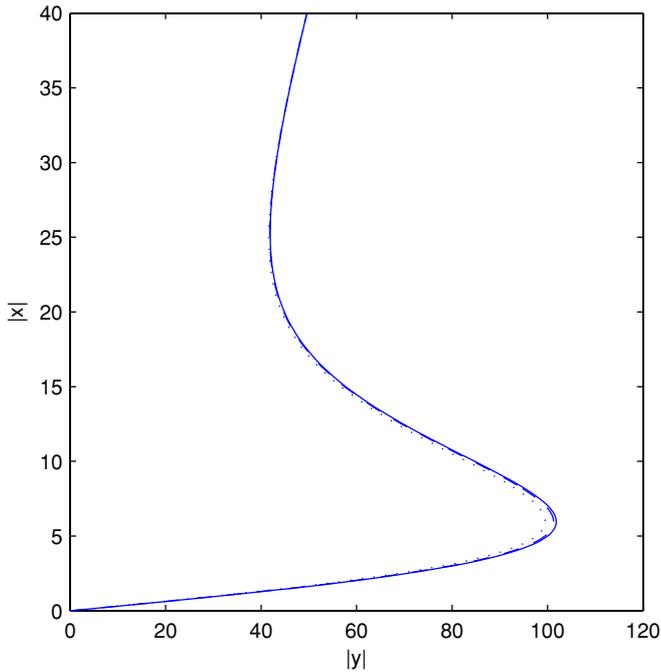

**Figure 8.** OB behaviour of the system for $\Delta_c = -4.1\gamma$, $\delta = 4.1\gamma$. Selected parameters are $\gamma_3 = \gamma_2 = \gamma$, $\Omega_c = 10\gamma$, $\Delta_p = 0$, $C = 400$, and $\eta = 0$ (solid), 0.5 (dashed), 0.95 (dotted).

## 4. Conclusion

The effect of quantum interference due to spontaneous emission in a three-level V-type system on the optical properties of the system is investigated both under the two-photon resonance condition and beyond it. We discuss the absorption, dispersion, group index and the behaviour of OB in two regimes. It is shown that the phase sensitivity of the absorption, the dispersion, the group index and OB will be valid only under the two-photon resonance condition. Beyond the two-photon resonance condition the phase sensitivity of the medium will disappear.